\documentclass[aps,prd,twocolumn,superscriptaddress,nofootinbib]{revtex4-2}
\usepackage[utf8]{inputenc}
\usepackage{amsbsy}
\usepackage{amsmath}
\usepackage{amsfonts}
\usepackage{amssymb}
\usepackage{xcolor}
\usepackage{physics}
\usepackage{graphicx}
\usepackage{mathrsfs}
\usepackage{hyperref}
\usepackage{cleveref}
\usepackage{float}

\begin{document}

\title{Quantum interferometric probe of neutron--hidden neutron oscillations}

\author{A. Capolupo}
\email{capolupo@sa.infn.it}
\affiliation{Dipartimento di Fisica ``E.R. Caianiello'', Università di Salerno,
	and INFN -- Gruppo Collegato di Salerno, Via Giovanni Paolo II, 132,
	84084 Fisciano (SA), Italy}

\author{G. Pisacane}
\email{gpisacane@unisa.it}
\affiliation{Dipartimento di Fisica ``E.R. Caianiello'', Università di Salerno,
	and INFN -- Gruppo Collegato di Salerno, Via Giovanni Paolo II, 132,
	84084 Fisciano (SA), Italy}

\author{A. Quaranta}
\email{aniello.quaranta@unicam.it}
\affiliation{School of Science and Technology, University of Camerino, Via Madonna delle Carceri, Camerino, 62032, Italy}

\author{P. B\"{o}ni }
\email{peter.boeni@tum.de}
\affiliation{Technical University of Munich, TUM School of Natural Sciences, Physics Department, 85748 Garching, Germany}

\begin{abstract}

The nature of dark matter remains an outstanding problem in particle physics and cosmology. Hidden-sector extensions of the Standard Model predict a neutral partner of the neutron, whose weak mixing with ordinary neutrons induces oscillations between visible and dark baryonic states. We show that macroscopic quantum interferometry provides a direct and experimentally accessible probe of this phenomenon. In particular, a Mach--Zehnder interferometer with very cold neutrons converts neutron--hidden neutron oscillations into measurable phase-dependent intensity modulations. By combining controlled phase shifts with tunable magnetic fields and material potentials, the setup enables a resonant exploration of the hidden-sector parameter space.
We find that existing cold-neutron facilities can probe mixing amplitudes down to $\epsilon_{nn'} \sim 10^{-14}\,\mathrm{eV}$ for mass splittings $\delta m \sim 10^{-9}\,\mathrm{eV}$, accessing a previously unexplored region of parameter space relevant to baryonic dark matter scenarios. These results establish neutron interferometry as a precision laboratory tool for testing hidden-sector physics.

\end{abstract}

\maketitle

\section{Introduction}
The nature of the dark sector remains one of the central open problems in modern physics, closely intertwined with unresolved issues in particle physics, such as the strong CP problem and the origin of neutrino masses and mixing \cite{DS1,DS2,DS3,DS4,DS5,DS5-1,DS5-2,DS6,DS7,DS8,SCP1,SCP2,SCP3,N1,N2,N3,N4,N5,N6,N7,N8,N9,N10,N11,N12,N13}. A wide class of extensions of the Standard Model (SM), including axion and axion-like particle scenarios \cite{AX1,AX2,AX3,AX4,AX5,AX6,AX7}, as well as hidden-sector frameworks \cite{DDM1,DDM3,DDM5,DDM4,AN4,DDM2,BaseNL,DDM6,DDM7,DDM8}, has been proposed to account for dark matter and related phenomena.
  In a broad class of such models, neutral fermions can couple feebly to SM baryons, giving rise to mixing phenomena and oscillations between visible and dark states. In particular, mirror-matter realizations and related hidden-sector theories predict the existence of a partner of the neutron $n'$, whose mixing with the ordinary neutron induces transitions between visible and hidden baryonic sectors \cite{MM1,MM2,MM3,MM4,MM5,MM6,MM7}. This mechanism provides a direct laboratory pathway to probe dark matter candidates with baryonic couplings.
  Experimental searches for neutron--hidden neutron oscillations have so far relied primarily on ultracold-neutron (UCN) storage experiments and beam disappearance measurements, which have established important constraints on the mixing parameters \cite{11,13,14,15,26,27}. Complementary approaches, including reactor and beam-based experiments such as MURMUR, STEREO, and SNS, have further explored this scenario over a range of mass splittings and oscillation times \cite{17,18,19,28}. Despite these efforts, a substantial region of parameter space characterized by small mass splittings ($\delta m \lesssim \mathrm{neV}$) and weak mixing amplitudes ($\epsilon_{nn'} \lesssim 10^{-14}\,\mathrm{eV}$) remains experimentally inaccessible. This limitation arises from intrinsic constraints of existing techniques, including finite neutron lifetimes, material losses, and, crucially, the absence of phase-sensitive observables capable of discriminating genuine conversion processes from systematic effects.

In this work, we show that quantum interferometry provides a qualitatively new route to probe hidden baryonic sectors. By exploiting the coherent evolution of neutron beams, interferometric configurations enable direct sensitivity to both phase and amplitude modifications induced by mixing with a hidden state. In contrast to conventional disappearance experiments, where the signal is encoded solely in neutron loss, interferometry allows for a differential measurement based on controlled phase manipulation and resonant enhancement. This leads to distinct and experimentally accessible signatures, namely, a correlated modulation of the interference pattern and a resonant suppression of the total detected intensity, that cannot be reproduced by standard absorption or instrumental effects.
We propose a Mach--Zehnder interferometer operating with very cold neutrons (VCNs) as a precision spectroscopic probe of neutron--hidden neutron mixing. Building upon earlier interferometric concepts \cite{Ebisawa}, the present configuration is specifically optimized to maximize the neutron time of flight, thereby enhancing the transition probability, while remaining compatible with current neutron-optics technology. The setup combines tunable magnetic fields with a controllable material phase shifter, enabling a resonant scan of the hidden-sector parameter space with high mass selectivity.

Our analysis demonstrates that existing cold-neutron facilities can achieve sensitivity to mixing amplitudes as small as $\epsilon_{nn'} \sim 10^{-14}\,\mathrm{eV}$ and mass splittings $\delta m \sim 10^{-9}\,\mathrm{eV}$, corresponding to oscillation times $\tau_{nn'} \sim 0.1\,\mathrm{s}$. This sensitivity probes a previously unexplored region of parameter space and is complementary to, and in part exceeds, the reach of current UCN storage and beam-based experiments, particularly in the regime of small mass splittings where interferometric resonance can be exploited.
These results establish neutron interferometry as a new precision frontier for laboratory searches of hidden baryonic sectors, demonstrating that quantum coherence can be harnessed as a powerful tool to access feeble mixing phenomena. More broadly, they open a realistic and scalable pathway to test dark-sector physics with table-top experiments.

\section{Theoretical Background: neutron - hidden neutron mixing}

We now outline the phenomenology of neutron mixing with a generic hidden state. While initially motivated by mirror matter models, where the dark sector is governed by a gauge group identical to the Standard Model, this formalism applies universally to any hidden sector containing a neutral fermion $n'$ (the ``hidden neutron") with mass $m_{n'}$. Dynamical mechanisms within this sector, such as spontaneous symmetry breaking, can generate a small mass splitting $\delta m = m_{n'} - m_n$ relative to the ordinary neutron mass $m_n$. Concurrently, a generic portal interaction drives a universal mixing between the ordinary and hidden neutral hadrons, ultimately leading to quantum oscillations between the two sectors.
The mixing of ordinary neutrons $n$ with their hidden counterpart $n'$ is analogous to two-flavor neutrino oscillations and admits a formally identical description. The physical ``flavor" states $n, n'$ are related to the mass states $n_1, n_2$ via
$
	\begin{pmatrix}
		n \\
		n'
	\end{pmatrix}
	=
	R(\theta)
	\begin{pmatrix}
		n_1 \\
		n_2
	\end{pmatrix},
$
where $\theta$ is the mixing angle and $R(\theta)$ is the rotation matrix. For vanishing momentum, the evolution equation in the flavor basis is described by the Schr\"{o}dinger equation
$
	i \frac{d}{d t}
	\begin{pmatrix}
		n \\
		n'
	\end{pmatrix}
	=
	H
	\begin{pmatrix}
		n \\
		n'
	\end{pmatrix},
$
with the effective mass Hamiltonian
$
	H = \begin{pmatrix}
		m_n + \Delta E & \epsilon_{n n'} \\
		\epsilon_{n n'} & m_n + \delta m
	\end{pmatrix},
	\label{mHamiltonian}
$
where $m_n$ is the neutron mass, $\epsilon_{n n'}$ is the mixing amplitude, $\Delta E$ is the energy contributed to the ordinary neutron by eventual external potentials (e.g., magnetic fields), and $\delta m $ is the mass splitting that survives in the limit $\Delta E = 0$ and $\epsilon_{n n'} \to 0$. Diagonalizing the Hamiltonian $H$, we obtain the mixing angle:
$
	\tan\left( 2 \theta \right) = \frac{2 \epsilon_{n n'}}{\Delta E - \delta m},
$
and the energy eigenvalues:
\begin{equation}
	m_{1,2} = \frac{1}{2} \left( 2 m_n + \Delta E + \delta m \pm \sqrt{\left( \Delta E - \delta m \right)^2 + 4 \epsilon_{n n'}^2} \right).
\end{equation}
For a generic $3$-momentum $\vec{k}$, the relativistic energy for a free particle of mass $m_j$ is $\omega_j = \sqrt{m_j^2 + k^2}$ with $k=|\vec{k}|$. Then the eigenstates of $H$, including kinetic energy, evolve according to $\ket{n_j (t)} = e^{- i \omega_j t} \ket{n_j}$, where $\ket{n_j} = \ket{n_j (t=0)}$.
A key feature of the system is the resonance condition
$
\delta m \simeq  \Delta E =\mu_n B + V,
$
for which the mixing is maximized. This condition provides a direct experimental handle to scan the hidden-sector parameter space by tuning the external magnetic field.

In the following, we exploit this resonant enhancement within a neutron interferometric setup, where the controlled phase evolution of the neutron beam allows for a precision measurement of the mixing parameters.

\section{Experimental setup}
\begin{figure}[ht]
		\centering
		\includegraphics[width=0.48\textwidth]{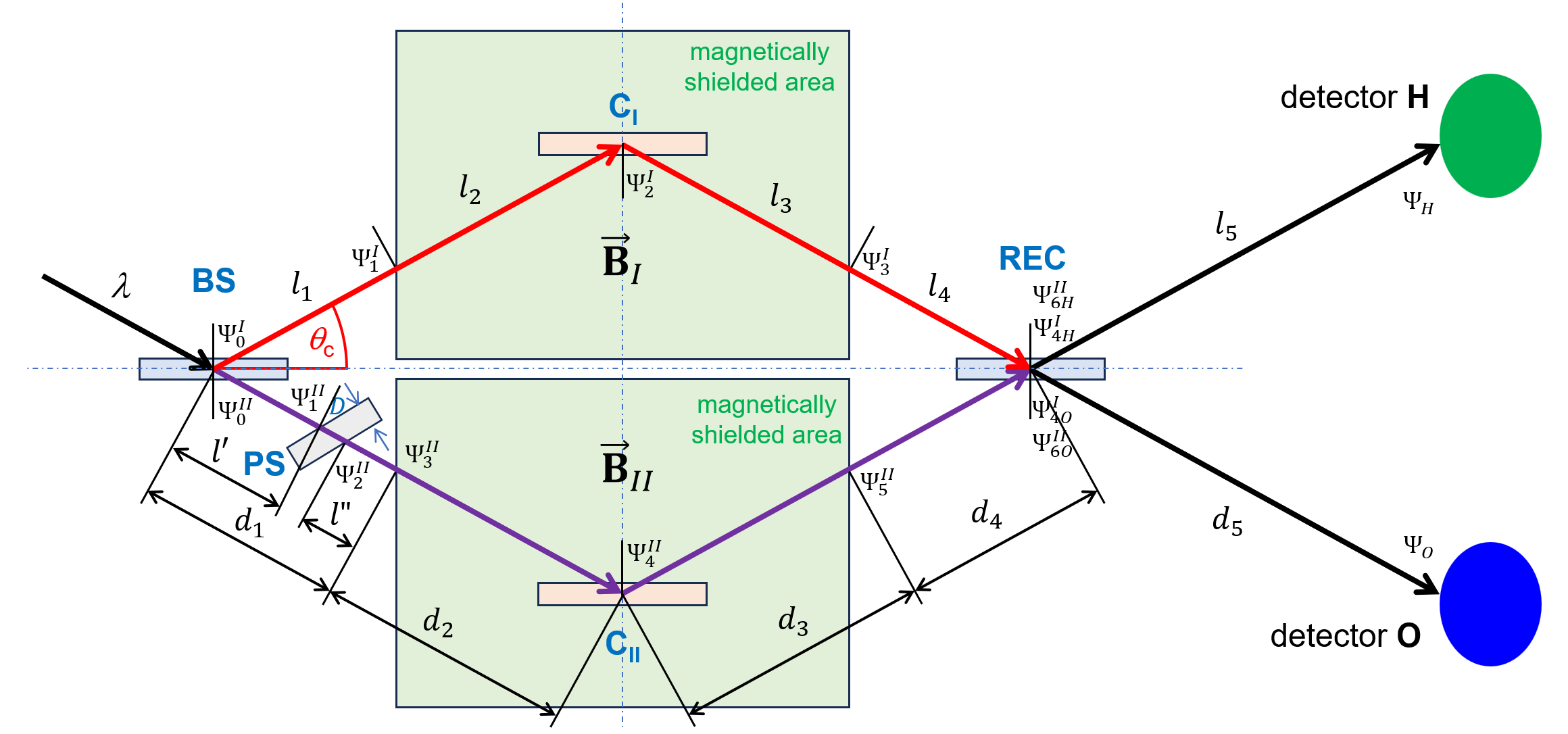}
		\caption{Scheme of the VCN interferometer, with path I (red arrows) and path II (violet arrows). The regions in green are magnetically shielded against disturbing external fields by mu-metal. The incident polarized beam is split by the beam splitter BS into two beams and, after their respective reflection from mirrors $C_{ I}$ and $C_{ II}$, recombined in the recombination mirror REC. The transmitted and reflected beams thereof are measured with detectors O and H. By varying the active magnetic field $\vec {\bf B}_{II}$, the system scans the mass splitting parameter space. For more details see the text.}
		\label{schema_V1}
	\end{figure}

In order to reveal the existence of hidden neutrons, and to measure the mass difference $\delta m$ and the mixing strength $\epsilon_{nn'}$, we propose using a macroscopic Mach-Zehnder-type neutron interferometer operating in the VCN regime, as sketched in Fig.~\ref{schema_V1}. While an initial concept based on cold neutrons was proposed by Ebisawa et al. \cite{Ebisawa}, our design is fundamentally optimized for VCNs to maximize the time of flight and, consequently, the transition probability. We employ half-transparent Ni-mirrors for the beam splitter (BS) and the recombining mirror (REC). The mirrors $C_{ I}$ and $C_{ II}$ reflect the neutrons onto REC where the beams interfere. The intensity of the reflected and transmitted beams is recorded with detectors O and H, respectively.

The angle of reflection of the neutrons at the mirror positions in degrees is given by $\theta_c = 0.099 m \lambda$, where $m$ is defined to be 1 for a Ni coating and $\lambda$ designates the wavelength of the neutrons in units of \AA\ \cite{Kolevatov2019}.
The transition probability of the $n \to n'$ oscillation scales quadratically with the time spent in the active magnetic region ($P_{n \to n'} \propto t^2$). Therefore, to maximize the time of flight and drastically enhance the signal, we propose using an incident VCN  beam
  with a   bandwidth $\delta_B = \Delta \lambda/\lambda \simeq 10\%-20\%$ centered at $\lambda = 40$ \AA\  (corresponding to a velocity $v \simeq 100$ m/s). It can be defined by means of a velocity selector that is installed upstream of the polarizing mirror. This broad acceptance is deliberately chosen to be compatible with the high-intensity VCN fluxes anticipated at next-generation facilities. At such long wavelengths, standard mirrors (e.g., $m = 1$) yield large reflection angles of $\theta_c \simeq 4^\circ$ (see appendix D). This enables a compact yet highly dispersive macroscopic geometry with flight paths in the active magnetic region of the order of $L \simeq 600$ mm.

    For the reflecting mirrors $C_{ I}$ and $C_{ II}$, Ni-mirrors providing a very high reflectivity of $99.9\%$ will be used, while for the BS and REC mirrors, a Ni-coating with a thickness $d = 122$ \AA\ will be tailored such that a reflectivity $R \simeq 50\%$ and a transmission of $T \simeq 50\%$ are obtained. To improve the transmission, the Si-substrate is thinned down at the beam area to approximately $50\,\mu$m. The incident beam is polarized using a polarizing mirror yielding a high polarization $P \simeq 99\%$ \cite{Schanzer2016}.

To achieve the required geometric stability, the polarizing and reflecting mirrors will be rigidly mounted between precisely machined glass base and cover plates, utilizing advanced manufacturing techniques established for standard neutron guides. Furthermore, the integration of kinematic adjustment mechanisms, akin to those used in ultrastable laser reference cavities, will guarantee exceptional alignment precision.  These design choices ensure that the interferometer operates as a truly monolithic neutron optical component, effectively minimizing the need for continuous active mechanical tuning during data acquisition.

The BS mirror directs the incident polarized neutron beam to the two arms designated as Path I (red arrows) and Path II (violet arrows) in Fig.~\ref{schema_V1}:

- Path I: The reflected neutron beam travels along a section $l_1$, where a magnetic guide field $B_g = 2$ mT is applied to maintain the polarization. Then it travels along the sections $l_2$ and $l_3$. In our baseline configuration, the magnetic field $B_I$ in this active region is kept at zero. Finally, it travels along sections $l_4$ and $l_5$, where a guide field $B_g = 2$ mT is applied.

- Path II: The transmitted neutron beam passes first through a region where a guide field $B_g = 2$ mT is applied. It contains the phase shifter made from a highly uniform Silicon wafer with a carefully calibrated thickness $D = 200\ \mu$m. It then completes the paths $d_2$ and $d_3$ (symmetric to $l_2$ and $l_3$), where the tunable active magnetic field $B_{II}$ is applied. Finally, the neutrons travel along sections $d_4$ and $d_5$, where again a guide field $B_g = 2$ mT is applied. The structural phase shift of the neutrons can be fine-tuned by a slight rotation of the Si block.

The active magnetic field $B_{II}$ is applied perpendicular to the neutron flight paths, i.e., perpendicular to the scheme shown in Fig.~\ref{schema_V1} and therefore parallel to the surface of the mirrors. Thus, the polarization of the neutrons is always kept parallel to the overall polarization $\pmb{P}$ of the neutron beam. By tuning the amplitude of $B_{II}$ (e.g., around 16.6 mT to probe $\delta m \sim 1$ neV), the setup acts as a precise spectrometer, actively scanning the hidden-sector parameter space.

\emph{Geometry.} Let us now specify the geometric details of the setup of Fig.~\ref{schema_V1}: The wavefronts of the beams are perpendicular to the plane of the interferometer. The angles of incidence and reflection are equal to $3^\circ$ with respect to the line joining the beam splitter BS and the recombination mirror REC (horizontal dashed line in Fig.~\ref{schema_V1}). We choose a highly optimized set of parameters tailored for VCN, which is reported in Table \ref{table}.

In this configuration, the flight paths in the non-active guide regions are $l:= l_1 = l_4 = d_1 = d_4 = 200 \ \mathrm{mm}$. The active magnetic region, accommodating the resonance scanning, has a total length $L = l_2 + l_3 = 600 \ \mathrm{mm}$ (where each section is $ l_2 = l_3 = 300 \ \mathrm{mm}$ in Path I). In Path II, to perfectly balance the interferometer, the corresponding lengths are $l' = l'' = 0.0999 \ \mathrm{m}$, leaving exactly $D = 200 \ \mu\mathrm{m}$ for the uniform silicon phase shifter. This yields an overall path length of $2l + L = 1000 \ \mathrm{mm}$ between BS and REC, maximizing the time of flight while maintaining a compact monolithic structure. In addition, we fix the distance to the detectors as $d:=l_5=d_5 = 500 \ \mathrm{mm}$.

The flight paths $l_2,\ l_3,\ d_2$ and $d_3$ are heavily shielded against external fields (for example, the Earth's magnetic field) using mu-metal to ensure a precise definition of the highly homogeneous fields along the active flight paths \cite{MuPAD}. Given the narrow width of the resonance, accessing $\epsilon_{nn'} \sim 10^{-14}\ \mathrm{eV}$ requires a relative field homogeneity of $\Delta B/B \sim 10^{-5}$ over the macroscopic drift length, a requirement strictly within the capabilities of state-of-the-art nuclear magnetic resonance (NMR) magnet technologies. To improve the accuracy of the overall field configuration, the guide fields $B_g = 2 \ \mathrm{mT}$ before and after the central active region are also shielded with mu-metal.

\emph{Phase shifter.} A slab of phase-shifting material with a thickness $D$ of refractive material, for example, Si, with a nuclear potential $V$, induces a time lag for beams traveling along Path II. For a plane wave traversing the slab, the phase lag is
$
\chi = k D_{eff}(n_r - 1)
$
\cite{Lemmel2010}, where $n_r$ is the refractive index given by
$
n_r = \frac{K}{k} = \sqrt{1 - \frac{V - V_{\text{air}}}{E - V_{\text{air}}}},
$
with $E = \frac{k^2 \hbar^2}{2m_n} + V_{\text{air}} = \frac{K^2 \hbar^2}{2m_n} + V$ being the total energy of the neutron. Here, $D_{eff}$ is the effective length of the beam path of the neutrons in the slab, which can be fine-tuned by adjusting the rotation angle of the Si wafer, and $V$ is the slab potential defined as
$
V = \frac{2\pi \hbar^2 N b_c}{m_n}.
$
Here, $b_c$ is the coherent scattering length, $N$ is the atomic density, and $m_n$ is the neutron mass. The potential of air, $V_{\text{air}}$, can be neglected in most cases, although it depends slightly on temperature, pressure, and humidity. In our baseline setup, we employ a highly uniform monocrystalline Si slab with a nominal thickness $D = 200 \ \mu\mathrm{m}$. To prevent spurious spatial smearing of the interference fringes across the beam profile, the slab must exhibit a total thickness variation well below the effective wavelength inside the medium. Commercial ultra-flat silicon wafers achieved through advanced chemomechanical polishing easily meet this sub-micron thickness tolerance.

\section{Hamiltonian and background elements}

We first consider, for simplicity, a monochromatic beam of VCNs with momentum $k \ll m_{n}$ (setting $\hbar = c = 1$) and negligible momentum spread $\delta_B$. The discussion for a Gaussian wave packet is reported in Appendix C. Such a beam is well described by a plane wave. For a specific value of $m_n$, the kinetic energy is:
$
\frac{K^2(V)}{2m_n} = \frac{k^2}{2m_n} \left(1-\frac{V}{E} \right) .
$
Here, we have assumed $V_{air} = 0$. The physical states $\ket{n}$ and $\ket{n'}$ evolve according to the effective mass Hamiltonian $M(V,B)$, obtained from  $H$, by setting $\Delta E = + \mu_n B + V$, i.e.,
\begin{equation}
M(V,B) = \begin{pmatrix}
	m_n + \mu_n B + V & \epsilon_{n n'} \\ \epsilon_{n n'} & m_n + \delta m
\end{pmatrix}.
\end{equation}
Note that, in Path I, $B = B_g$ along the sections $l_1$ and $l_4$, and $B = B_I$ along the sections $l_2$ and $l_3$. Moreover, in Path II, $B = B_g$ along $d_1$ and $d_4$, and $B = B_{II}$ along $d_2$ and $d_3$ (see Fig. \ref{schema_V1}). The full Hamiltonian, comprising the kinetic term, is
\begin{equation}
H(V,B) = M(V,B) + \frac{k^2}{2M(V,B)}\left(1- \frac{V}{E} \right),
\end{equation}
where the slight abuse of notation $1/M$ is to be understood in terms of the eigenvalues of $M$. We introduce the mixing angle
$
\theta(V,B)  = \frac{1}{2} \arctan \left(\frac{2 \epsilon_{n n'}}{ + \mu_n B + V-\delta m} \right),
$
the eigenstates $\ket{n_1(V,B)}, \ket{n_2(V,B)}$, and the corresponding eigenvalues
\begin{eqnarray}\nonumber
	\Omega_{1,2}(V,B)  &=& \frac{1}{2}\Big(2m_n + \delta m + \mu_n B
	\\ &+& V \pm \sqrt{(\mu_n B +V  -\delta m )^2 + 4\epsilon_{n n'}^2}  \Big),
\end{eqnarray}
and we set
$
E_{1,2}(V,B) = \Omega_{1,2}(V,B) + \frac{k^2}{2\Omega_{1,2}(V,B)} \left(1-\frac{V}{E} \right)
$
for the eigenvalues of the complete Hamiltonian. The ordinary - hidden basis is related to the eigenstates $\ket{n_1(V,B)}$ and $\ket{n_2(V,B)}$ through
\begin{eqnarray}\label{Rotation1}
	\nonumber \ket{n} &=& \cos \theta(V,B) \ket{n_1(V,B)} + \sin \theta(V,B) \ket{n_2(V,B)} \\
	\nonumber \ket{n'} &=& - \sin \theta(V,B) \ket{n_1(V,B)} + \cos \theta(V,B) \ket{n_2(V,B)}.
	\\
\end{eqnarray}
The group velocity is defined for each of the eigenvalues as $v_{1,2}(V,B) = \frac{K(V)}{\Omega_{1,2}(V,B)}$. If the neutron had a definite mass (no mixing), we would expect a group velocity $v = v_n(V) \simeq \frac{K(V)}{m_n}$. The difference among all these velocities is, however, negligible \footnote{In fact, one has:
	\begin{eqnarray}\nonumber
		K(V)\left(\frac{1}{\Omega_{1,2}(V,B)} - \frac{1}{m_n}\right) = \frac{K(V)}{m_n}\left[ \frac{- (\mu_n B + V + \delta m)}{2\Omega_{1,2}(V,B)}\right.
		\\\nonumber
		\left.\mp \frac{\sqrt{(\mu_n B + V - \delta m)^2 + 4\epsilon_{n n'}^2}}{2\Omega_{1,2}(V,B)}\right].
	\end{eqnarray}
	The term in brackets goes as $\frac{\delta E}{m_n}$, where $\delta E$ is the sum of various small (compared to $m_n$ and even to $k$) energy shifts.}. Therefore, we consider a single velocity in the beam given by $v = v_n(V)$.

The final elements impacting the beam propagation are the mirrors, which are characterized by a transmission and reflection coefficient $T$ and $R$, respectively, related to the neutron fraction which is transmitted $|T|^2$ and reflected $|R|^2$. As explained in Appendix \ref{appendix D}, in the following it is save to set $T^2 = 1$ and $R^2 = 1$ because the absorption of the BS and REC mirrors is  close to 0. Therefore, the static attenuation can be factored out by the magnetic lock-in extraction. Finally, we assume that the mirrors are fully transparent for the hidden neutron component $n'$.

\section{ Propagation of the beam}

We now have all the elements needed to describe the evolution of the beams along paths I and II. At each step, we   express the state of the beam on the neutron - hidden neutron basis, as
\begin{equation}
 \ket{\psi^{J}(t)} = N^{J}(t) \ket{n} + M^{J}(t) \ket{n'}  ,
\end{equation}
where $N^{J}(t)$ and $M^{J}(t)$ are the amplitudes for the neutron and the hidden-neutron respectively.

 - Beam I:
 We assume that the neutron beam impinges the beam splitter BS at $t=0$. It is comprised only of a neutron component because the hidden part is not reflected by the polarizer.  Let $N_0^{I}$ and $N_0^{II}$ denote the initial amplitudes for beam I and II (see appendix A).\footnote{$N_0^{I}$ and $N_0^{II}$   include  the term $\sqrt{1-A_{BS}}$, with $A_{BS}$ designating the absorption coefficient in the beam splitter.} The initial state for beam I is therefore
 $
  \ket{\psi^I(0)} = N_0^I \ket{n}.
$
The beam traverses a region with a small guide field of 2 mT of length $l_1 =l $. One has the normalized state reported in appendix A. Then it traverses the area filled with magnetic field $B_I$ for a distance $l_2 =L $, up to the central mirror $C_I$.
The state at the entrance of the magnetic area has the form $
  \ket{\psi^I_1} = N_1^I \ket{n}+ M_1^I \ket{n'}
$.
Since the hidden neutron component is entirely transmitted through the mirror $C_I$, the state after reflection from $C_I$ is just
$
  \ket{\psi^I_2} = R_C^I  N_2^I \ket{n}
$ with $N_2^I$ being an appropriate combination of $N_1^I $ and $M_1^I $ (see also appendix A).

We now iterate the same process for another length
$L$.
At the exit of the magnetic area, the state is
$\ket{\psi_3^{I}}=  N_3^{I} \ket{n} + M_3^{I} \ket{n'}$.

The beam traverses another  field region with 2 mT of a length $l_4 = l$, thus, after the recombination mirror, one has
$
 \ket{\psi^I_{4H}}  =\sqrt{1-A_{REC}} R_{REC}N^I_4 \ket{n} $ and
 $\ket{\psi^I_{4O}}  = \sqrt{1-A_{REC}}T_{REC}N^I_4 \ket{n} + M^I_4 \ket{n'}
$,
respectively, directed towards the H and O detector. Note that the absorption of the mirrors $A_{REC}\simeq 0$ (see appendix D), therefore it can be  neglected. All the mathematical details and the explicit expressions of the coefficients $N^I_j$,  $M^I_j$, (with $j=0,...,4$) $R_C^I$, $R_{REC}$ and $T_{REC}$ are reported in appendices A and B.

 - Beam II:
The initial state is now given by $\ket{\psi_0^{II}} = N_0^{II}  \ket{n}+ M_0\ket{n'}$. The presence of a hidden neutron component is due to the lack of interaction between hidden and ordinary matter, and therefore with the $BS$. This results in complete transmission of the hidden part by the devices. Then the beam evolves in a region with a magnetic guide field of intensity 2 mT for a distance $l'$. At the entrance of the slab, the state is given by $\ket{\psi_1^{II}}= N_1^{II} \ket{n} + M_1^{II}\ket{n'}$.
After crossing the slab for length $D_{eff}$, the state is $\ket{\psi_2^{II}}= N_2^{II} \ket{n} + M_2^{II}\ket{n'}$. Next, the beam traverses a region, with the magnetic guide field of 2 mT, for a length $l''$. The wave function at the entrance of the shielded region is denoted by $\ket{\psi_3^{II}}=  N_3^{II} \ket{n} + M_3^{II} \ket{n'}$.
Only the neutron part is reflected by the central mirror $C_{II}$, so that the state, immediately after $C_{II}$ is  $\ket{\psi_4^{II}}= R^{II}_C N_4^{II} \ket{n}$.
Subsequently, the beam exits the area with magnetic field $B_{II}$ with state $\ket{\psi_5^{II}}= N_5^{II}\ket{n} + M_5^{II} \ket{n'}$
and evolves in a region with 2 mT up to the recombination mirror. After the recombination mirror, one has
$ \ket{\psi_{6O}^{II}}=R_{REC}N_6^{II}\ket{n}$, and
 $ \ket{\psi_{6H}^{II}}= T_{REC}N_6^{II}\ket{n} + M^{II}_6 \ket{n'} .$

Then, the $H$ and $O$ beams, directed towards the respective detectors, are described by the following amplitudes:
\begin{eqnarray}\nonumber
 \ket{\psi^{in}_H} = (T_{REC}N_6^{II}+R_{REC}N_4^{I})\ket{n} +  M^{II}_6  \ket{n'} \\
\nonumber \ket{\psi^{in}_O} = (T_{REC}N_4^{I}+R_{REC}N_6^{II})\ket{n} +  M^{I}_4  \ket{n'}.
 \\
\end{eqnarray}
Finally, both of them evolve in a free region of length $l_5 = d_5$ before detection. At the detectors, dropping the $n'$ part, one has  $  \ket{\psi_H} = (N^I_H +N_H^{II})\ket{n}
$  and $\ket{\psi_O} = (N^I_O +N_O^{II}) \ket{n} \,$,
thus, the intensities are given by
\begin{eqnarray} \label{Ieq8}
 I_{O,H} &\propto& \langle \psi_{O,H} \ket{\psi_{O,H}}
\\ & = & |N^I_{O,H}|^2 + |N^{II}_{O,H}|^2 + 2 \mathrm{Re} \left(\bar{N}^I_{O,H} N^{II}_{O,H} \right), \nonumber
\end{eqnarray}
where we denote with $\bar{N}$ the complex conjugate of $N$. These output intensities are sensitive to the presence of neutron - hidden neutron mixing.

\section{Numerical analysis}
In the light of the parametrization for the reflection and transmission coefficients
$R_J = \cos \Gamma_J e^{-i\phi_{RJ}}$ and $ T_J = \sin \Gamma_J e^{-i\phi_{TJ}}$,
we set the various parameters as outlined in Appendix B and reported in Table \ref{table}.

\begin{table}[h]
	\centering
	\renewcommand{\arraystretch}{1.1}
	\footnotesize
	\begin{tabular}{|c|c|}
		\hline
		\textbf{Parameters} & \textbf{Value} \\
		\hline
		$\varphi_{T_{BS}} = \varphi_{T_C^I} = \varphi_{T_C^{II}}=\varphi_{T_{REC}}$ & 0 \\
		\hline
		$\varphi_{R_{BS}} = \varphi_{R_C^{I}} = \varphi_{R_C^{II}} = \varphi_{R_{REC}} $ & $\frac{\pi}{2}$ \\
		\hline
		$L_P$ & 0.2 m\\
		\hline
	    $l_1 = l_4 = d_1 = d_4 := l$ & 0.2 m \\
		\hline
		$l_2 = l_3 = d_2 = d_3 := L$ & 0.3 m \\
		\hline
		$l'=l''$ & 0.0999 m \\
		\hline
		$D$  & $200 \ \mu$m \\
		\hline
		$l_5 = d_5 := d$ & 0.5 m \\
		\hline
		$\lambda$ & 40 \AA \\
		\hline
		$m$ & 1 \\
		\hline
		$\Gamma_{R_C^{II}} = \Gamma_{R_C^{I}}$ & $0$ \\
		\hline
		$\Gamma_{REC}$ & $\frac{\pi}{4}$\\
		\hline
		$V$  & 54 neV \\
		\hline
	\end{tabular}
	\caption{Baseline parameters of the VCN interferometer. For our simulations, we assume the use of mirror with $m = 1$. $L_P$ denotes the distance between the polarizing mirror and $BS$.}
	\label{table}
\end{table}

Our analysis is based on the experimental setup proposed in Fig. \ref{schema_V1} operating with the parameters given in Table \ref{table}. This macroscopic VCN configuration amplifies the neutron time-of-flight in the active magnetic region, pushing the sensitivity to mixing strengths of the order of $\epsilon_{n n'} \sim \mathcal{O}(10^{-14}) \ \mathrm{eV}$, corresponding to an oscillation time $\tau_{n n'} \sim \mathcal{O}(10^{-1}) \ \mathrm{s}$, where $\tau_{n n'}=1/\epsilon_{n n'}$.
In the following, we set $B_I = B_{II} = B$. In this case, the net phase difference between the paths is driven strictly by the silicon phase shifter.   By tuning $B$, the resonance condition $\delta m \simeq \mu_n B + V$ can be achieved for different values of $\delta m$ and $\epsilon_{nn'}$.

\begin{figure}[h!]
	\centering
	\begin{minipage}{0.48\textwidth}
		\includegraphics[width=\linewidth]{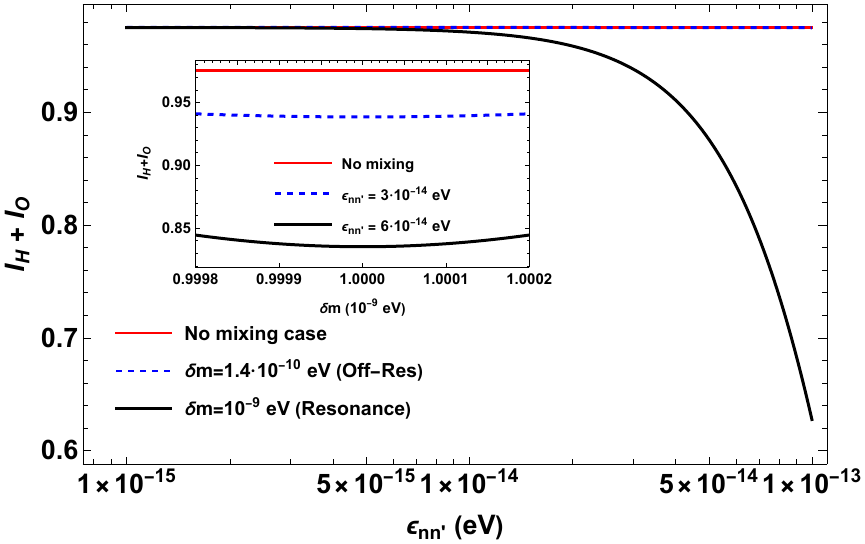}
		\par\noindent
	\end{minipage}
	\caption{(Color online) Main panel:    total intensity $I_{tot} = I_H + I_O$ as a function of the coupling strength $\epsilon_{n n'}$ for fixed values of the mass splitting $\delta m$. Inset: $I_{tot}$ as a function of $\delta m$ for fixed values of $\epsilon_{n n'}$. The suppression of the intensity appears at the resonance condition at $\delta m =  10^{-9} \ \mathrm{eV}$ in the beam II. The magnetic fields are set to $B_I = B_{II} \approx 16.58 \ \mathrm{mT}$.}
	\label{fig:1}
\end{figure}

Figure~\ref{fig:1} shows the dependence of the total intensity $I_{\mathrm{tot}} = I_H + I_O$ on the mixing parameters $\delta m$ and $\epsilon_{nn'}$. For a   mass splitting of $1 \ \mathrm{neV}$, the resonance is sharply localized around $B \simeq 16.58 \ \mathrm{mT}$. The red solid line indicates the no-mixing case. A clear suppression of $I_{\mathrm{tot}}$ is observed at resonance.
Although $I_{\mathrm{tot}} < 1$ due to an intrinsic $\sim 2\%$ absorption in the phase shifter (PS), the additional reduction at resonance provides a direct signature of neutron disappearance into the hidden sector.

\begin{figure}[h!]
	\centering
	\includegraphics[width=0.48\textwidth]{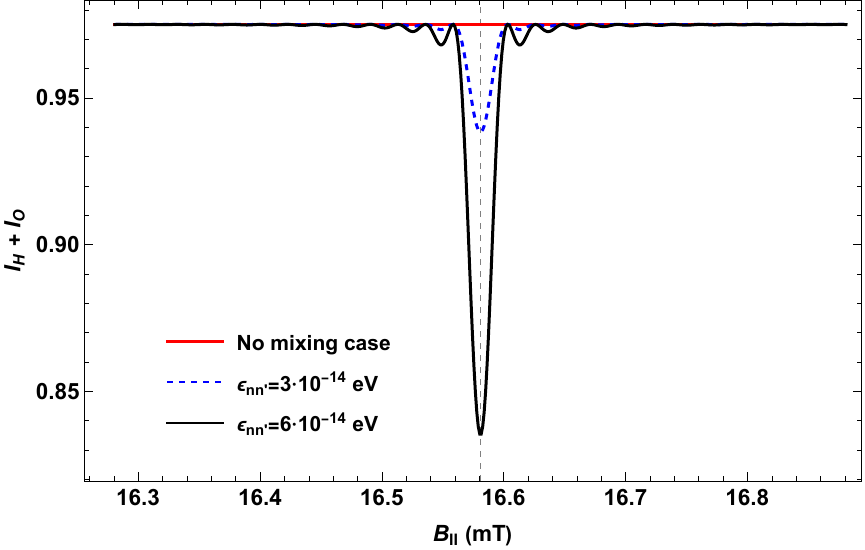}
	\caption{(Color online) Total intensity $I_{tot} = I_H + I_O$ for a monochromatic beam as a function of the applied magnetic fields $B_I = B_{II}$, assuming a fixed mass splitting $\delta m = 10^{-9} \ \mathrm{eV}$.}
	\label{fig:2}
\end{figure}

\begin{figure}[h!]
	\centering
	\includegraphics[width=0.48\textwidth]{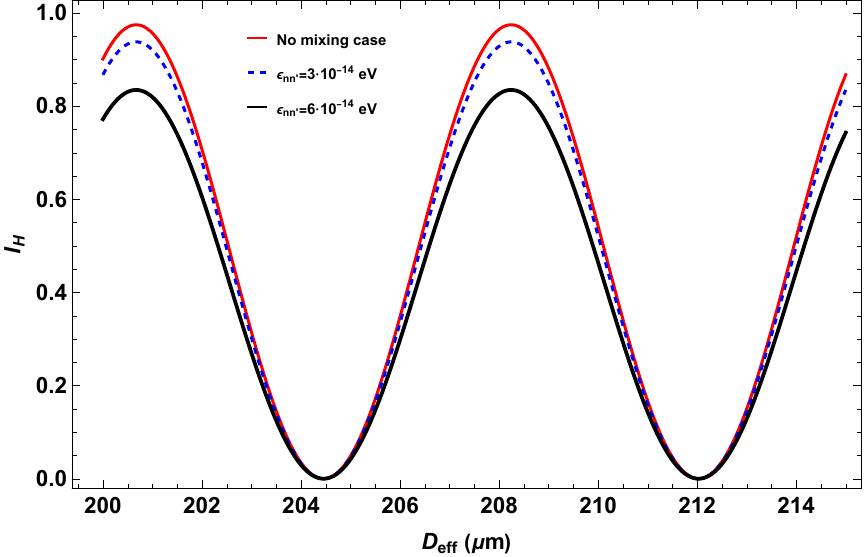}
	\caption{(Color online) Plot of $I_H$ as a function   $D_{eff}$. The red solid line represents the standard expected fringes without mixing. The blue dotted and black solid curves show the suppressed interference pattern under resonance conditions ($\delta m = 10^{-9} \ \mathrm{eV}$, $B \approx 16.58 \ \mathrm{mT}$) for   $\epsilon_{nn'} = 3 \cdot 10^{-14} \ \mathrm{eV}$ and $\epsilon_{nn'} = 6 \cdot 10^{-14} \ \mathrm{eV}$, respectively.}
	\label{fig:3}
\end{figure}

Figure~\ref{fig:2} displays the dependence of $I_{tot}$ on the applied magnetic field. The pronounced selectivity of the signal imposes stringent requirements on the spatial homogeneity of the field, thereby motivating the adoption of a dedicated mu-metal shielding scheme along the active paths.
The interferometric response is encoded in the modulation of the transmitted intensity as a function of the effective thickness $D_{eff}$ of the silicon phase shifter. Figure~\ref{fig:3} shows $I_H$ as a function  of $D_{eff}$. At resonance, neutron conversion into the hidden sector produces a clear deviation from the no-mixing scenario.
To prevent spurious path-length mismatches during the variation of the effective thickness $D_{eff}$, the monolithic design structurally constrains Path II to perfectly match Path I ($d_1 = l_1 = 0.2\ \mathrm{m}$). This is achieved by permanently fixing the symmetric drift regions $l'$ and $l''$ to exactly compensate for the Si block thickness $D$. The magnetic fields are fixed to the values adopted in Figs.~\ref{fig:1} and \ref{fig:2}.

\begin{figure}[h!]
	\centering
	\includegraphics[width=0.5\textwidth]{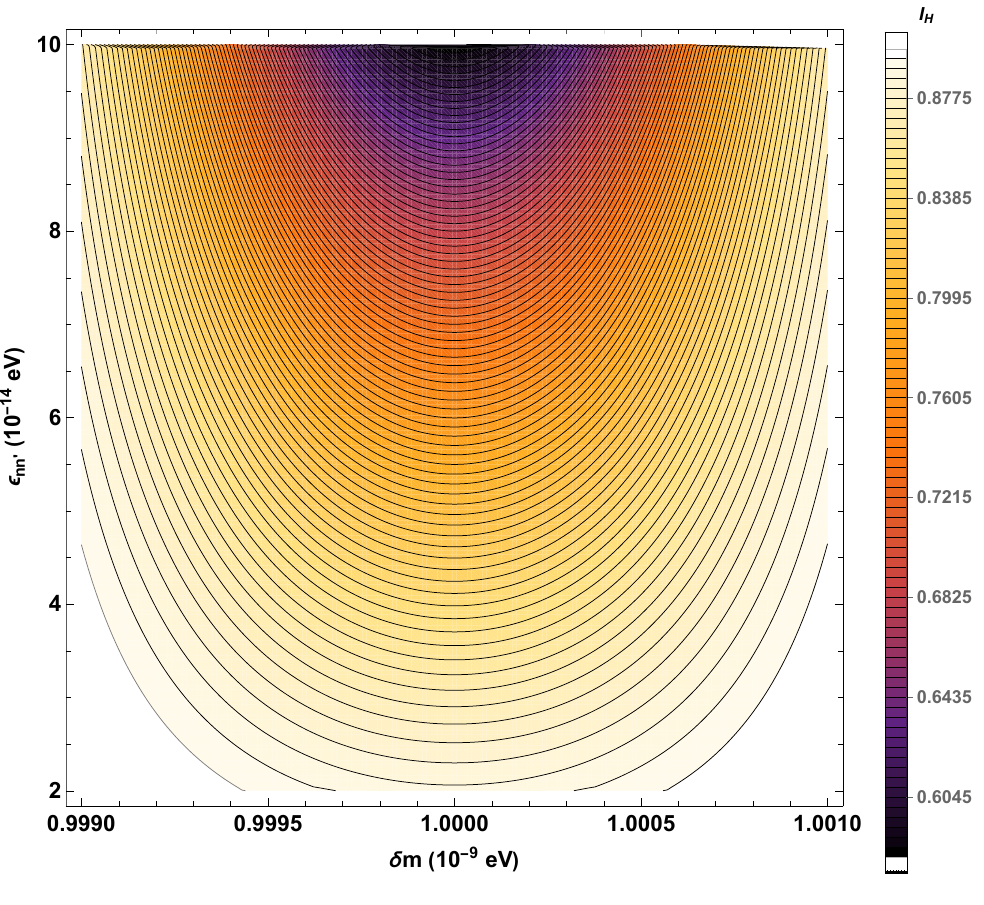}
    \caption{(Color online) Contour plot   $I_H$ as a function   $\delta m$ and   $\epsilon_{n n'}$. The dark vertical band represents the resonance condition at $\delta m = 1.0 \cdot 10^{-9} \ \mathrm{eV}$, achieved by setting   $B  \approx 16.58 \ \mathrm{mT}$.}
	\label{fig:4}
\end{figure}

The overall sensitivity of the experimental configuration is summarized in Fig.~\ref{fig:4}, which presents a contour map of the transmitted intensity $I_H$ in the $(\delta m, \epsilon_{nn'})$ parameter space.  The numerical analysis indicates that the VCN setup can probe mixing amplitudes as small as $\epsilon_{nn'} \sim 10^{-14}\ \mathrm{eV}$.

\begin{figure}[h!]
	\centering
	\includegraphics[width=0.5\textwidth]{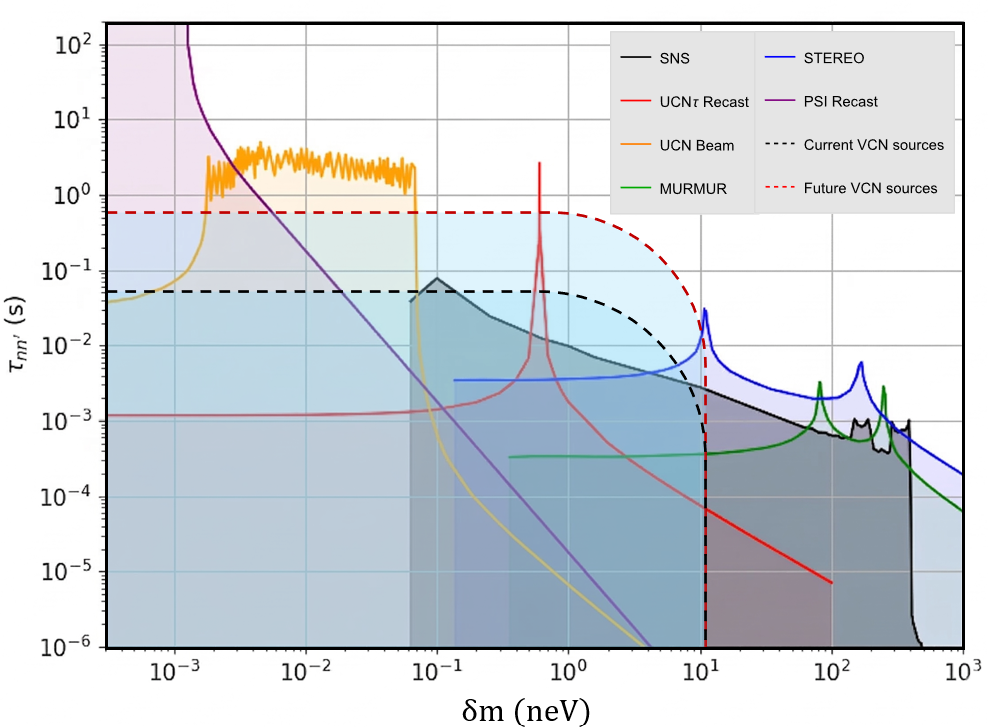}
\caption{(Color online)
 Exclusion plot for  $\tau_{nn'}$ as a function of   $\delta m$ and comparison with the existing constraints. Sky blue shaded region: parameter space accessible to the proposed VCN interferometer. The lower boundary corresponds to the baseline sensitivity  at current high-flux continuous sources, The upper boundary indicates the projected reach ($\tau_{nn'} \sim 1 \ \mathrm{s}$), achievable at next-generation pulsed facilities such as ESS. }
 \label{confronto}
\end{figure}

Fig.~\ref{confronto} shows the exclusion plot for the neutron–hidden neutron oscillation time $\tau_{nn'}$ as a function of the mass splitting $\delta m$. The sky blue-shaded area delineates the previously unexplored parameter space rendered accessible by the proposed VCN interferometer.

To properly contextualize this reach, it is necessary to analyze the physical limitations of existing bounds. Current limits derived from UCN storage \cite{11,13,14,15,26} and UCN beam disappearance experiments \cite{27} are fundamentally constrained by finite neutron lifetimes, wall-collision losses, and the lack of a phase-sensitive observable. Similarly, bounds from reactor and spallation-based experiments like STEREO \cite{18}, MURMUR \cite{17}, and SNS \cite{19,28} rely on overall flux depletion or are optimized for different kinematic regimes.

In contrast, the VCN interferometer explicitly targets the ultra-low mass splitting region ($\delta m \sim 10^{-9}\ \mathrm{eV}$) by exploiting the resonant matching condition $\delta m \simeq \mu_n B + V$. As indicated by the shape of the exclusion region, this resonant enhancement, combined with the differential phase extraction provided by the Mach-Zehnder geometry, allows the VCN probe to drastically bypass current experimental walls, pushing the sensitivity to much feebler mixing amplitudes.

The boundaries of the sky blue region reflect the impact of the neutron source on the statistical reach. The lower boundary corresponds to the baseline sensitivity ($\tau_{nn'} \sim 0.1\ \mathrm{s}$) attainable with standard integration times at current high-flux continuous sources. The upper boundary demonstrates the ultimate projected reach of $\tau_{nn'} \sim 1\ \mathrm{s}$. This maximum sensitivity is expected at next-generation pulsed facilities, such as the European Spallation Source (ESS) \cite{Andersen-2020}, where the combination of unprecedented peak fluxes and time-of-flight background suppression will allow for a significantly cleaner extraction of the disappearance signal.

\emph{- Systematic uncertainties and the Magnetic Lock-in Technique: -}
 A key challenge in neutron interferometry is the discrimination of genuine conversion losses ($n \to n'$) from systematic effects, such as material absorption and structural misalignments. The proposed scheme addresses this issue via an intrinsic magnetic modulation technique. Since standard absorption processes are insensitive to the applied field $B$, periodic switching between an on-resonance configuration ($B_{\mathrm{ON}}$) and a detuned off-resonance setting ($B_{\mathrm{OFF}} = B_{\mathrm{ON}} + \Delta B$) enables the extraction of a differential signal $(I_{\mathrm{OFF}} - I_{\mathrm{ON}})/I_{\mathrm{OFF}}$. This procedure suppresses static optical losses, allowing for the detection of missing neutron fractions at the $0.1\%$ level, in agreement with the sensitivity inferred from the numerical analysis.

\emph{  - Gaussian wave packet: - }
Until now, we have considered monochromatic plane waves. In order to increase the signal statistics, one may use a beam with a broad bandwidth of the order $\delta_B = 10\% - 20\%$. This corresponds to the typical high-flux spectrum available for Very Cold Neutrons (VCNs) at advanced sources. The bandwidth $\delta_B$ can be defined by means of a velocity selector or a bandpass multilayer. We consider a Gaussian wave packet, which for mathematical convenience is defined by discrete momentum values $k_n = 2\pi n/a$. The details are given in Appendix C. The evolution of the wave packet follows straightforwardly from the plane wave analysis performed above.
In the numerical analysis for the wave packet, we consider a central wavelength $\lambda_0 = 40\,\text{\AA}$, a spatial parameter $a = 16000\,\text{\AA}$, and the baseline parameters reported in Table \ref{table}.

Although the standard interference contrast (the difference $I_H^G - I_O^G$) is reduced compared to the monochromatic case due to the short longitudinal coherence length ($l_c = \lambda_0^2 / \Delta\lambda$) of the broadband packet washing out the spatial fringes, the $n \to n'$ conversion acts as an absorptive effect. Consequently, the drop in the total intensity $I_{tot}^G$ at the resonance condition is preserved, independent of the longitudinal coherence length of the packet. Numerical analysis confirms that the experimental setup is sensitive to the hidden sector parameter space even with broadband VCN beams.

\begin{figure}[h!]
	\centering
	\begin{minipage}{0.48\textwidth}
		\centering
		\includegraphics[width=\linewidth]{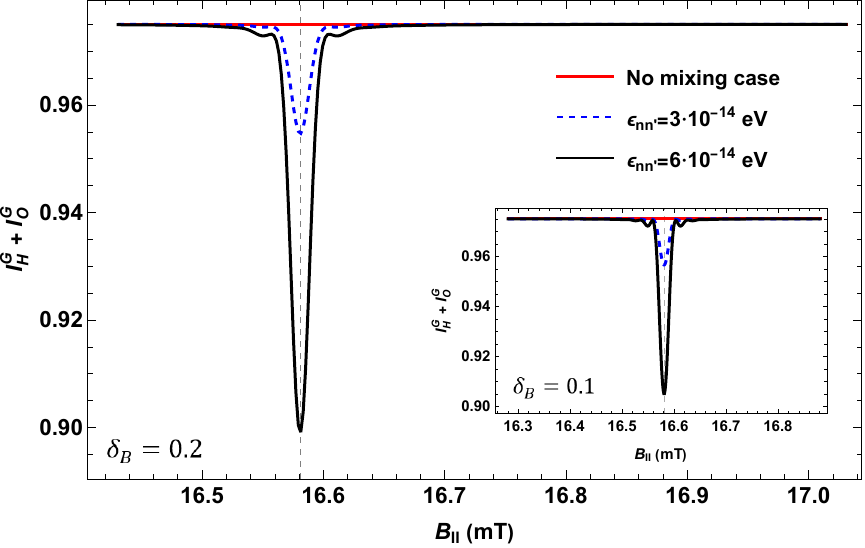}
	\end{minipage}
	\caption{(Color online) Total intensity $I_{tot}^{G}$ for a Gaussian wave packet as a function of the applied magnetic field $B=B_I = B_{II}$, assuming a fixed  $\delta m = 10^{-9} \ \mathrm{eV}$. Main panel:  $\delta_B = 0.1$ ($10\%$ bandwidth). Inset: $\delta_B = 0.2$ ($20\%$ bandwidth). The resonance drop is preserved in both polychromatic beam scenarios.}
	\label{fig:5}
\end{figure}

Figure \ref{fig:5} displays the total intensity $I_{tot}^{G}$ for a wave packet as a function of the magnetic field $B $, with momentum spreads of $\delta_B = 0.1$ and $\delta_B = 0.2$. The retention of the signal drop indicates that the magnetic modulation technique remains effective despite beam chromaticity, allowing for high-statistics measurements.

\section{Estimation of sensitivity}

We now estimate the expected signal strength at the detectors of the interferometer. If the source brilliance is known, the neutron intensity $I$ entering the interferometer can be written as \cite{boeni2014}
$
	I = A \cdot \Delta\lambda \cdot \Omega \cdot \Psi,
$
where $A$, $\Delta\lambda$, and $\Omega$ denote the beam size, wavelength bandwidth, and solid angle at the entrance of the apparatus. For the VCN beamline at the Institut Laue-Langevin (ILL PF2) \cite{yellowbook1986}, the quoted spectral flux, combined with the beamline geometry and optical acceptance, corresponds to an effective brilliance of order $\Psi \sim 10^{5}\ \mathrm{cm^{-2}\,s^{-1}\,\AA^{-1}\,sr^{-1}},$ which we use as a benchmark estimate for the unmonochromatized beam.
Adopting representative experimental parameters, $A = 10 \times 10 \ \mathrm{mm}^2$, $\Delta\lambda = 4 \ \mathrm{\AA}$ (corresponding to a relative bandwidth $\delta_B \simeq 10\%$ at $\lambda = 40 \ \mathrm{\AA}$), and $\Omega = 10^{-4} \ \mathrm{sr}$, we obtain an incoming intensity of order $I \sim 60 \ \mathrm{s^{-1}}$. Accounting for polarization selection at the first mirror and a typical reflectivity of $\sim 99\%$ for the subsequent optical elements, the expected count rate at the detectors is reduced to $\mathcal{O}(20)\ \mathrm{s^{-1}}$.

This rate, while modest compared with standard cold-neutron experiments, is sufficient for precision measurements over extended integration times. At $20\ \mathrm{s^{-1}}$, the apparatus accumulates approximately $1.7\times10^6$ events per day, corresponding to a Poisson uncertainty of order $\sqrt{N} \simeq 1.3\times10^3$.

In the presence of neutron--hidden neutron mixing, the observable effect is a reduction in the detected neutron flux governed by the conversion probability $P_{n\to n'}$, which depends on the magnetic-field configuration and the neutron propagation time. For representative parameter choices leading to an average conversion probability at the percent level, the expected deficit is of order $10^4$ events per day, i.e. well above the purely statistical fluctuations. This indicates that the effect should be statistically resolvable with sufficient integration time, provided that instrumental systematics are controlled at the same level.

A key aspect of the measurement is the control of systematic effects. Ordinary neutron losses due to absorption in the interferometer optics (e.g. REC and BS mirrors) are expected to be not relevant (see Appendix D), comparable to the anticipated signal. To separate these contributions, the experiment can be operated in a differential mode by alternating between magnetic-field configurations corresponding to on-resonance and off-resonance conditions. Since material absorption is insensitive to the applied magnetic field, it contributes as a common-mode effect, while any correlated modulation of the detected intensity can be attributed to the mixing-induced conversion probability, up to residual field-dependent instrumental effects that can be calibrated independently.

Finally, operation at next-generation pulsed facilities such as the European Spallation Source (ESS) \cite{Andersen-2020} would provide a distinct experimental advantage beyond raw intensity. By exploiting the time structure of the pulsed beam, one could apply time-of-flight (TOF) techniques to drastically suppress uncorrelated background noise. This substantial background reduction, intrinsic to the temporal beam structure, would significantly enhance the signal-to-noise ratio, extending the reach of the experiment toward smaller mixing probabilities and longer oscillation times.

\section{Conclusions}

We have demonstrated that macroscopic quantum interferometry provides a powerful and experimentally accessible method to probe neutron--hidden neutron mixing. By exploiting the coherent evolution of very cold neutron beams in a Mach--Zehnder configuration, the proposed setup enables a phase-sensitive and differential measurement of hidden-sector effects, going beyond the capabilities of conventional disappearance experiments.
The key result of this work is the identification of a robust and distinctive observable: a correlated modification of the interference pattern accompanied by a resonant suppression of the total detected intensity. This combined signature arises from the coherent interplay between phase evolution and conversion into the hidden sector, and cannot be mimicked by standard absorption or instrumental effects. The use of magnetic tuning further allows for a spectroscopic scan of the hidden-sector parameter space with high selectivity.
Our quantitative analysis shows that existing cold-neutron facilities can achieve sensitivity to mixing amplitudes as small as $\epsilon_{nn'} \sim 10^{-14}\,\mathrm{eV}$ and mass splittings $\delta m \sim 10^{-9}\,\mathrm{eV}$, corresponding to oscillation times $\tau_{nn'} \sim 0.1\,\mathrm{s}$. This regime is complementary to, and in part extends beyond, the reach of current ultracold-neutron storage and beam-based experiments, particularly in the domain of small mass splittings where resonant enhancement can be fully exploited. The feasibility of the measurement is supported by realistic estimates of neutron fluxes, counting statistics, and systematic control through magnetic modulation techniques.

These results establish neutron interferometry as a new precision frontier for laboratory searches of hidden baryonic sectors. More broadly, they demonstrate that quantum coherence can be harnessed as a sensitive probe of feeble mixing phenomena, opening a scalable and conceptually new pathway to test dark-sector physics with table-top experiments. Future implementations at next-generation neutron sources are expected to further extend the accessible parameter space, providing a competitive and complementary approach to uncover physics beyond the Standard Model.

\vspace{5mm}

\section*{Acknowledgments}
Partial financial support from MUR and INFN is acknowledged. A.C. also acknowledges the COST Action CA1511 Cosmology and Astrophysics Network for Theoretical Advances and Training Actions (CANTATA).

\appendix

\section{Propagation of the beam, explicit computations}
We consider a polarized neutron beam that propagates from the polarizing mirror to the beam splitter over a distance of length $L_P$ before reaching it at time $t=t_0$. Assuming that the beam emerging from the polarizer at $t=t_{-1}$ contains only neutrons, the initial state can be written as $\ket{\psi(t_{-1})} = e^{i\phi}\ket{n}$. The state at the beam splitter is then

\begin{widetext}
\begin{eqnarray} \nonumber \ket{\psi(L_P)}&=&e^{i\phi} \left[\cos \theta (0,0) e^{\frac{-iE_1(0,0)L_P}{v_n}}\ket{n_1(0,0)} + \sin \theta (0,0) e^{\frac{-iE_2(0,0)L_P}{v_n}}\ket{n_2(0,0)} \right] \\ \nonumber &=& e^{i\phi} \left[\cos \theta (0,0) e^{\frac{-iE_1(0,0)L_P}{v_n}} \left(\cos \theta (0,0) \ket{n} - \sin \theta (0,0) \ket{n'} \right) \right. \\ \nonumber &+& \left.\sin \theta (0,0) e^{\frac{-iE_2(0,0)L_P}{v_n}}\left(\sin \theta (0,0) \ket{n} + \cos \theta (0,0) \ket{n'} \right) \right] \\ \nonumber &=& e^{i\phi}\left[\cos^2 \theta (0,0)e^{\frac{-iE_1(0,0)L_P}{v_n}} + \sin^2 \theta (0,0) e^{\frac{-iE_2(0,0)L_P}{v_n}} \right] \ket{n}\\ &+&e^{i\phi} \cos\theta(0,0) \sin \theta(0,0) \left[e^{\frac{-iE_2(0,0)L_P}{v_n}} - e^{\frac{-iE_1(0,0)L_P}{v_n}} \right] \ket{n'}, \end{eqnarray}

Here $\theta(0,0)$, $E_J(0,0)$, and $\ket{n_i(0,0)}$ denote the mixing angle, eigenenergies, and eigenstates evaluated for $V=0$ and $B=0$, and the rotation defined by Eq.~\eqref{Rotation1} (and its inverse) has been used.

At the beam splitter, the state can be expressed as
\begin{equation}
	\ket{\psi(0)} = N_0\,\ket{n} + M_0\,\ket{n'},
\end{equation}
where the coefficients are
\begin{align} N_{0} &= e^{i\phi} \left[ \cos^{2} \theta (0,0)\, e^{ \frac{- i\, E_{1}(0,0)\,L_P}{v_n}} + \sin^{2} \theta (0,0)\, e^{ \frac{- i\, E_{2}(0,0)\,L_P}{v_n}} \right], \\[4mm] M_{0} &= e^{i\phi}\, \cos \theta (0,0)\, \sin \theta (0,0)\left[ e^{\frac{- i\, E_{2}(0,0) L_P}{v_n}} - e^{\frac{- i\, E_{1}(0,0)L_P}{v_n}} \right]. \end{align}

Beam I: We assume that the neutron beam impinges the beam splitter at $t=0$. It is comprised only of a neutron component because the hidden part is not reflected by the $BS$. Let $N_0^{I}$ and $N_0^{II}$ denote the initial amplitudes for beam I and II. Note that any neutron absorption by $BS$ is encoded within $N_0^I$ and $N_0^{II}$ that are given in terms of $N_0$ and $M_0$ as:

\begin{eqnarray}
N^I_0&=&\sqrt{1-A_{BS}} R_{BS} N_0\\
N^{II}_0&=&\sqrt{1-A_{BS}} T_{BS} N_0
\end{eqnarray}

The initial state for beam I is therefore
 $
  \ket{\psi^I(0)} = N_0^I \ket{n} .
$
The beam traverses a region with a guide field $B_g = 2$ mT and with zero potential of length $l_1 =l$.\footnote{We point out that both beam paths are exposed to $B_g = 2$ mT, thus the phase shift due to $B_g$ is canceled.}
From Eq.\eqref{Rotation1}, we have the following state normalized to the initial amplitude:

\begin{eqnarray}
 \nonumber \frac{\ket{\psi^{I}(l)}}{N_0^I}&=& \cos \theta (0,0) e^{\frac{-iE_1(0,0)l}{v_n}}\ket{n_1(0,0)} + \sin \theta (0,0) e^{\frac{-iE_2(0,0)l}{v_n}}\ket{n_2(0,0)}  \\ \nonumber
 &=& \cos \theta (0,0) e^{\frac{-iE_1(0,0)l}{v_n}} \left(\cos \theta (0,0) \ket{n} - \sin \theta (0,0) \ket{n'} \right) \\
 \nonumber  &+& \sin \theta (0,0) e^{\frac{-iE_2(0,0)l}{v_n}}\left(\sin \theta (0,0) \ket{n} + \cos \theta (0,0) \ket{n'} \right)  \\ \nonumber &=& \left[\cos^2 \theta (0,0)e^{\frac{-iE_1(0,0)l}{v_n}} + \sin^2 \theta (0,0) e^{\frac{-iE_2(0,0)l}{v_n}} \right]\ket{n}+\left[\cos\theta(0,0) \sin \theta(0,0) e^{\frac{-iE_2(0,0)l}{v_n}} - e^{\frac{-iE_1(0,0)l}{v_n}} \right]\ket{n'}. \\ &&
\end{eqnarray}
We can read off the amplitudes at the entrance of the magnetic area as
\begin{eqnarray}
 N^{I}_1 &=& N_0^I \left[\cos^2 \theta (0,0)e^{\frac{-iE_1(0,0)l}{v_n}} + \sin^2 \theta (0,0) e^{\frac{-iE_2(0,0)l}{v_n}} \right] \\
 M^{I}_1 &=& N_0^I \cos\theta(0,0) \sin \theta(0,0) \left[e^{\frac{-iE_2(0,0)l}{v_n}} - e^{\frac{-iE_1(0,0)l}{v_n}} \right].
\end{eqnarray}
Beam I does then traverses the area filled with magnetic field $B_I$ for a distance $l_2 = L$, up to the central mirror $C_I$. The state at the entrance is  now
$
 \ket{\psi^I_1} = N^{I}_1 \ket{n} + M^{I}_1 \ket{n'}  .
$
We proceed in the same way as before, just we have to  consider the presence of the magnetic field $B_I$. Since the hidden neutron component is entirely transmitted by the mirror $C_I$,  the state after reflection from $C_I$ is of the form $R^I_C N^I_2 \ket{n}$. At the distance $L$, immediately before $C_I$, the neutron and hidden neutron states are (note that now $V=0$ and $B=B_I$):
\begin{eqnarray*}
 \ket{n} &\rightarrow& \cos \theta (0,B_I) e^{\frac{-iE_1(0,B_I)L}{v_n}} \ket{n_1(0,B_I)} + \sin \theta (0,B_I) e^{\frac{-iE_2(0,B_I)L}{v_n}} \ket{n_2(0,B_I)} \\
 &=& \cos \theta (0,B_I) e^{\frac{-iE_1(0,B_I)L}{v_n}}\left( \cos \theta (0,B_I) \ket{n} - \sin \theta (0,B_I) \ket{n'}\right) \\
 &+& \sin \theta (0,B_I) e^{\frac{-iE_2(0,B_I)L}{v_n}} \left(\sin \theta (0,B_I) \ket{n} + \cos \theta (0,B_I) \ket{n'} \right) \\
 &=& \left[\cos^2 \theta (0,B_I) e^{\frac{-iE_1(0,B_I)L}{v_n}} + \sin^2 \theta (0,B_I) e^{\frac{-iE_2(0,B_I)L}{v_n}}\right] \ket{n} \\ &+& \cos \theta(0,B_I) \sin \theta (0,B_I) \left[e^{\frac{-iE_2(0,B_I)L}{v_n}} -  e^{\frac{-iE_1(0,B_I)L}{v_n}}\right] \ket{n'}
\end{eqnarray*}
and
\begin{eqnarray*}
 \ket{n'} &\rightarrow& \cos \theta (0,B_I) e^{\frac{-iE_2(0,B_I)L}{v_n}} \ket{n_2(0,B_I)} - \sin \theta (0,B_I) e^{\frac{-iE_1(0,B_I)L}{v_n}} \ket{n_1(0,B_I)} \\
 &=& \cos \theta (0,B_I) e^{\frac{-iE_2(0,B_I)L}{v_n}}\left( \cos \theta (0,B_I) \ket{n'} + \sin \theta (0,B_I) \ket{n}\right) \\
 &-& \sin \theta (0,B_I) e^{\frac{-iE_1(0,B_I)L}{v_n}} \left(-\sin \theta (0,B_I) \ket{n'} + \cos \theta (0,B_I) \ket{n} \right) \\
 &=& \left[\cos^2 \theta (0,B_I) e^{\frac{-iE_2(0,B_I)L}{v_n}} + \sin^2 \theta (0,B_I) e^{\frac{-iE_1(0,B_I)L}{v_n}}\right] \ket{n'} \\ &+& \cos \theta(0,B_I) \sin \theta (0,B_I) \left[e^{\frac{-iE_2(0,B_I)L}{v_n}} -  e^{\frac{-iE_1(0,B_I)L}{v_n}}\right] \ket{n} ,
\end{eqnarray*}
respectively, so that we have
 \begin{eqnarray}
  \nonumber N_2^I &=& N_1^I\left[\cos^2 \theta (0,B_I) e^{\frac{-iE_1(0,B_I)L}{v_n}} + \sin^2 \theta (0,B_I) e^{\frac{-iE_2(0,B_I)L}{v_n}}\right] \\
  &+& M_1^I \cos \theta(0,B_I) \sin \theta (0,B_I) \left[e^{\frac{-iE_2(0,B_I)L}{v_n}} -  e^{\frac{-iE_1(0,B_I)L}{v_n}}\right]
 \end{eqnarray}
and the state, immediately after $C_I$ is just
$
 \ket{\psi_2^I} = R^I_C N_2^I \ket{n} .
$
 We now iterate the same process for another length $L$. From now on the relations among the intermediate amplitudes shall be omitted and reported in the appendix B. At the exit of the magnetic area the state is
$
 \ket{\psi_3^I} = N^I_3 \ket{n} + M^I_3 \ket{n'}.
$
The beam traverses another field region with $B_g = 2$ mT of a length $l_4=l$, thus,   after the recombination mirror,  one has
$
 \ket{\psi^I_{4H}}  = \sqrt{1-A_{REC}}R_{REC}N^I_4 \ket{n} $ and
 $\ket{\psi^I_{4O}}  =\sqrt{1-A_{REC}} T_{REC}N^I_4 \ket{n} + M^I_4 \ket{n'}
$,
respectively, directed towards the H and O detector.

\section{Relations among intermediate amplitudes and transmission and reflection coefficients}

Here, we provide explicitly all the intermediate amplitudes. For the beam $I$:
\begin{eqnarray}\nonumber
 N^I_{3} = R^I_C N^I_2\left[\cos^2 \theta (0,B_I) e^{\frac{-iE_1(0,B_I)L}{v_n}} + \sin^2 \theta (0,B_I) e^{\frac{-iE_2(0,B_I)L}{v_n}}\right]
 \\ \label{b1}
 M^I_3 =R^I_CN^I_2 \cos \theta(0,B_I) \sin \theta (0,B_I) \left[e^{\frac{-iE_2(0,B_I)L}{v_n}} - e^{\frac{-iE_1(0,B_I)L}{v_n}} \right]
\end{eqnarray}
\begin{eqnarray}\nonumber
 N^I_{4} &=& N^I_3  \left[\cos^2 \theta (0,0) e^{\frac{-iE_1(0,0)l}{v_n}} + \sin^2 \theta (0,0) e^{\frac{-iE_2(0,0)l}{v_n}}\right] + M^I_3 \cos \theta(0,0) \sin \theta (0,0) \left[ e^{\frac{-iE_2(0,0)l}{v_n}} - e^{\frac{-iE_1(0,0)l}{v_n}}\right]
 \\ \label{b2}
 M^I_{4} &=&  M^I_3  \left[\cos^2 \theta (0,0) e^{\frac{-iE_2(0,0)l}{v_n}} + \sin^2 \theta (0,0) e^{\frac{-iE_1(0,0)l}{v_n}}\right] +N^I_3 \cos \theta(0,0) \sin \theta (0,0) \left[ e^{\frac{-iE_2(0,0)l}{v_n}} - e^{\frac{-iE_1(0,0)l}{v_n}}\right] \ .
\end{eqnarray}
For the beam II, the  free evolution on the length $l'$ is characterized by
\begin{eqnarray}\nonumber
 N^{II}_{1} &=& N^{II}_0  \left[\cos^2 \theta (0,0) e^{\frac{-iE_1(0,0)l'}{v_n}} + \sin^2 \theta (0,0) e^{\frac{-iE_2(0,0)l'}{v_n}}\right]
 \\ \nonumber
 &+& M_0  \cos \theta(0,0) \sin \theta (0,0) \left[ e^{\frac{-iE_2(0,0)l'}{v_n}} - e^{\frac{-iE_1(0,0)l'}{v_n}}\right]
 \\ \nonumber
 M^{II}_{1} &=&  N^{II}_0 \cos \theta(0,0) \sin \theta (0,0) \left[ e^{\frac{-iE_2(0,0)l'}{v_n}} - e^{\frac{-iE_1(0,0)l'}{v_n}}\right]
 \\ \label{b3}
 &+&M_0\left[\cos^2 \theta (0,0) e^{\frac{-iE_2(0,0)l'}{v_n}} + \sin^2 \theta (0,0) e^{\frac{-iE_1(0,0)l'}{v_n}} \right].
\end{eqnarray}
The evolution within the slab, considering  the modified speed $v_n(V)$, gives the following amplitudes right after the slab:
\begin{eqnarray}\nonumber
 N^{II}_{2} &=& N^{II}_1  \left[\cos^2 \theta (V,0) e^{\frac{-iE_1(V,0)D_{eff}}{v_n(V)}} + \sin^2 \theta (V,0) e^{\frac{-iE_2(V,0)D_{eff}}{v_n(V)}}\right] \\
 \nonumber&+& M^{II}_1 \cos \theta(V,0) \sin \theta (V,0) \left[ e^{\frac{-iE_2(V,0)D_{eff}}{v_n(V)}} - e^{\frac{-iE_1(V,0)D_{eff}}{v_n(V)}}\right]
 \\ \nonumber
 M^{II}_{2} &=&  M^{II}_1  \left[\cos^2 \theta (V,0) e^{\frac{-iE_2(V,0)D_{eff}}{v_n(V)}} + \sin^2 \theta (V,0) e^{\frac{-iE_1(V,0)D_{eff}}{v_n(V)}}\right]\\
 &+&N^{II}_1 \cos \theta(V,0) \sin \theta (V,0) \left[ e^{\frac{-iE_2(V,0)D_{eff}}{v_n(V)}} - e^{\frac{-iE_1(V,0)D_{eff}}{v_n(V)}}\right]   .\label{b4}
\end{eqnarray}
After the slab, the beam traverse again a free region of length $l''$ ($l'+l''+D_{eff}=d_1$). At the entrance of the magnetic area the amplitudes are
\begin{eqnarray}\nonumber
 N^{II}_{3} &=& N^{II}_2  \left[\cos^2 \theta (0,0) e^{\frac{-iE_1(0,0)l''}{v_n}} + \sin^2 \theta (0,0) e^{\frac{-iE_2(0,0)l''}{v_n}}\right] + M^{II}_2 \cos \theta(0,0) \sin \theta (0,0) \left[ e^{\frac{-iE_2(0,0)l''}{v_n}} - e^{\frac{-iE_1(0,0)l''}{v_n}}\right]
 \\ \nonumber
 M^{II}_{3} &=&  M^{II}_2  \left[\cos^2 \theta (0,0) e^{\frac{-iE_2(0,0)l''}{v_n}} + \sin^2 \theta (0,0) e^{\frac{-iE_1(0,0)l''}{v_n}}\right] +N^{II}_2 \cos \theta(0,0) \sin \theta (0,0) \left[ e^{\frac{-iE_2(0,0)l''}{v_n}} - e^{\frac{-iE_1(0,0)l''}{v_n}}\right]   .
\end{eqnarray}
The neutron amplitude at $C_{II}$ is
\begin{eqnarray}\nonumber
 N^{II}_{4} &=& N^{II}_3  \left[\cos^2 \theta (0,B_{II}) e^{\frac{-iE_1(0,B_{II})L}{v_n}} + \sin^2 \theta (0,B_{II}) e^{\frac{-iE_2(0,B_{II})L}{v_n}}\right]
 \\ \label{b5}
  &+& M^{II}_3 \cos \theta(0,B_{II}) \sin \theta (0,B_{II}) \left[ e^{\frac{-iE_2(0,B_{II})L}{v_n}} - e^{\frac{-iE_1(0,B_{II})L}{v_n}}\right]
\end{eqnarray}
and immediately after, one has $\ket{\psi_{II}} = R_C^{II} N^{II} \ket{n}$ since the mirror part has been fully transmitted. At the exit of the magnetic area we have
\begin{eqnarray}\nonumber
 N^{II}_{5} &=& R^{II}_C N^{II}_4  \left[\cos^2 \theta (0,B_{II}) e^{\frac{-iE_1(0,B_{II})L}{v_n}} - \sin^2 \theta (0,B_{II}) e^{\frac{-iE_2(0,B_{II})L}{v_n}}\right]
 \\ \label{b6}
 M^{II}_{5} &=& R^{II}_C N^{II}_4 \cos \theta(0,B_{II}) \sin \theta (0,B_{II}) \left[ e^{\frac{-iE_2(0,B_{II})L}{v_n}} - e^{\frac{-iE_1(0,B_{II})L}{v_n}}\right] \ .
\end{eqnarray}
Finally
\begin{eqnarray}\nonumber
 N^{II}_{6} &=& N^{II}_5  \left[\cos^2 \theta (0,0) e^{\frac{-iE_1(0,0)l}{v_n}} + \sin^2 \theta (0,0) e^{\frac{-iE_2(0,0)l}{v_n}}\right] + M^{II}_5 \cos \theta(0,0) \sin \theta (0,0) \left[ e^{\frac{-iE_2(0,0)l}{v_n}} - e^{\frac{-iE_1(0,0)l}{v_n}}\right]
  \\ \nonumber
 M^{II}_{6} &=&  M^{II}_5  \left[\cos^2 \theta (0,0) e^{\frac{-iE_2(0,0)l}{v_n}} + \sin^2 \theta (0,0) e^{\frac{-iE_1(0,0)l}{v_n}}\right] +N^{II}_5 \cos \theta(0,0) \sin \theta (0,0) \left[ e^{\frac{-iE_2(0,0)l}{v_n}} - e^{\frac{-iE_1(0,0)l}{v_n}}\right]   .
 \\ \label{b7}
\end{eqnarray}
The amplitudes at the detectors are:
\begin{eqnarray}\nonumber
 N^{I}_{H} &=&\sqrt{1-A_{REC}} R_{REC}N^{I}_4  \left[\cos^2 \theta (0,0) e^{\frac{-iE_1(0,0)d}{v_n}} + \sin^2 \theta (0,0) e^{\frac{-iE_2(0,0)d}{v_n}}\right]
 \\ \nonumber
N^{II}_{H} &=& \sqrt{1-A_{REC}}T_{REC}N^{II}_6  \left[\cos^2 \theta (0,0) e^{\frac{-iE_1(0,0)d}{v_n}} + \sin^2 \theta (0,0) e^{\frac{-iE_2(0,0)d}{v_n}}\right]
\\ \nonumber
 &+& M^{II}_6 \cos \theta(0,0) \sin \theta (0,0) \left[ e^{\frac{-iE_2(0,0)d}{v_n}} - e^{\frac{-iE_1(0,0)d}{v_n}}\right]
 \\ \nonumber
N^{I}_{O} &=& \sqrt{1-A_{REC}}T_{REC}N^{I}_4  \left[\cos^2 \theta (0,0) e^{\frac{-iE_1(0,0)d}{v_n}} + \sin^2 \theta (0,0) e^{\frac{-iE_2(0,0)d}{v_n}}\right]
\\ \nonumber
&+& M^{I}_4 \cos \theta(0,0) \sin \theta (0,0) \left[ e^{\frac{-iE_2(0,0)d}{v_n}} - e^{\frac{-iE_1(0,0)d}{v_n}}\right]
\\ \label{b8}
 N^{II}_{O} &=& \sqrt{1-A_{REC}}R_{REC}N^{II}_6  \left[\cos^2 \theta (0,0) e^{\frac{-iE_1(0,0)d}{v_n}} + \sin^2 \theta (0,0) e^{\frac{-iE_2(0,0)d}{v_n}}\right] \ .
\end{eqnarray}

We set the transmission and reflection coefficients as follows:

\begin{align} \nonumber
	T_{BS}\left(\Gamma_{REC}, \varphi_{T_{BS}}\right) &= \sin\left(\Gamma_{BS}\right) e^{-i \varphi_{T_{BS}}},
	& R_{BS}\left(\Gamma_{BS}, \varphi_{R_{BS}}=\frac{\pi}{2}\right) &= \cos(\Gamma_{BS}) e^{-i \frac{\pi}{2}},
\\ \nonumber
	T_C^I(\Gamma_{R_C^{I}}, \varphi_{T_C^{I}} ) &= \sin(\Gamma_{R_C^{I}}) e^{-i \varphi_{T_C^{I}} },
	& R_C^I\left(\Gamma_{R_C^{I}}, \varphi_{R_C^{I}} =\frac{\pi}{2}\right) &= \cos(\Gamma_{R_C^{I}}) e^{-i \frac{\pi}{2} },
\\ \nonumber
	T_C^{II}(\Gamma_{R_C^{II}}, \varphi_{T_C^{II}}) &= \sin(\Gamma_{R_C^{II}}) e^{-i \varphi_{T_C^{II}}},
	& R_C^{II}\left(\Gamma_{R_C^{II}}, \varphi_{R_C^{II}}=\frac{\pi}{2}\right) &= \cos(\Gamma_{R_C^{II}}) e^{-i \frac{\pi}{2}},
 \\ \label{b9}
	T_{REC}\left(\Gamma_{REC}, \varphi_{T_{REC}}\right) &= \sin\left(\Gamma_{REC}\right) e^{-i \varphi_{T_{REC}}},
	& R_{REC}\left(\Gamma_{REC}, \varphi_{R_{REC}}=\frac{\pi}{2}\right) &= \cos(\Gamma_{REC}) e^{-i \frac{\pi}{2}}.
\end{align}

Note that the absorption coefficients in the central mirrors $A_{C}^{I}$, and $A_{C}^{II}$ are contained in the transmission coefficients $T_{C}^{I}$ and $T_{C}^{II}$, which quantify the fraction of neutrons that are not reflected. It is worth noting that we assigned a phase shift of $\pi/2$ to the reflection coefficients in Table~\ref{table}.
This phase relation is required to strictly satisfy the conservation of particle number.
As discussed by Zeilinger~\cite{Zeilinger1981}, the condition $|T|^2 + |R|^2 = 1$ is not sufficient to guarantee unitarity; for each mirror, the interference cross-terms must also vanish, satisfying the relation $R T^* + T R^* = 0$.
We applied this specific phase difference to ensure that the final intensities derived in Eq.~\eqref{Ieq8} satisfy the conservation condition $I_O + I_H \leq 1$ (in terms of the initial amplitude).

\section{Gaussian wave packet }

We take the initial state $\ket{\psi^I(0)}$ as a Gaussian wave packet, which, for mathematical convenience, we consider discrete and centered at the wave vector $k_0 = 2\pi/\lambda_0$ of the neutrons.\footnote{%
    In the expression for $\ket{\psi^I(0)}$ one should in principle consider an integral instead of a summation;
	the discretization is introduced for numerical convenience, as the integral cannot be performed  analytically. The continuum may be recovered in the $a \rightarrow 0$ limit.
	In any case, this approximation deviates negligibly from the exact result, as can be numerically verified.}

\begin{equation*}
	\ket{\psi^I(0)}
	= \left(\sum_{m=1}^{\infty} e^{-\frac{(k_m-k_0)^2}{2\sigma_k^2}}\right)^{-1/2}
	\sum_{n=1}^{\infty} e^{-\frac{(k_n-k_0)^2}{4\sigma_k^2}}
	\ket{\psi^I_{k_n}(0)}.
\end{equation*}
Here, the wave vectors are given by $k_n = \frac{2\pi n}{a}$, with a  finite lattice spacing $a$, $\sigma_{k}$ is the standard deviation in $k$-space, which can be parameterized as $\sigma_{k}\simeq \delta_B k_0$, where $k_0$ is  the central value of the Gaussian wave packet, and $\delta_B $ is  the fractional spread in momentum around $k_0$. Note that in the sum, $n$  assumes only positive values, because we are interested in wave packets propagating in the forward direction.
In practice, the range of $n$-values can be restricted so that only wave vectors $k_n \simeq k_0$ are included that provide meaningful contributions to the sums.
We assume that the reflection and transmission of the mirrors are only weakly dependent on the momentum of the neutrons in the range considered. Carrying out calculations similar to the ones  performed for the monochromatic case, we obtain the following normalized intensities $I_O$ and $I_H$:
\begin{equation*}
  I_O^\mathrm{G}
  \;=\;\left(\sum_{m=1}^{\infty} e^{-\frac{(k_m-k_0)^2}{2\sigma_k^2}}\right)^{-2}
  \sum_{n=1}^\infty
    e^{-\frac{(k_n-k_0)^2}{2\sigma_k^2}} \; I_O(k_n) \,,
\quad \quad \quad
  I_H^\mathrm{G}
 \;=\;\left(\sum_{m=1}^{\infty} e^{-\frac{(k_m-k_0)^2}{2\sigma_k^2}}\right)^{-2}
\sum_{n=1}^\infty
e^{-\frac{(k_n-k_0)^2}{2\sigma_k^2}} \; I_H(k_n) \, ,
\end{equation*}
where $I_O(k_n)$ and $I_H(k_n)$ denote the normalized intensities at $O$ and $H$ for a plane wave with momentum $k_n$.

\section{Properties of reflecting and half-transparent Ni mirrors}
\label{appendix D}

The optical elements of the interferometer are based on standard $m=1$ reflection coatings using elemental Ni deposited on silicon substrates. The set-up is designed for an angle of incidence $\theta = 3^\circ$ and neutrons with a wavelength $\lambda = 40$ \AA\ (Figure \ref{fig:Reflection-Transmission-BP}). Therefore, $\theta$ is well below the critical angle of reflection of Ni that is given by $\theta_c \simeq 0.099$ $(^\circ/$\AA) $\cdot$ 40 \AA\ $ = 3.96^\circ$. Note that $\theta = 3^\circ$ corresponds to a momentum transfer $Q = 0.0164$ \AA$^{-1}$. We have calculated the absorption and the reflectivity of the used mirrors by means of the matrix method \cite{Yamada-1978}.

\begin{figure}[htb]
	\centering
	\includegraphics[width=0.7\textwidth]{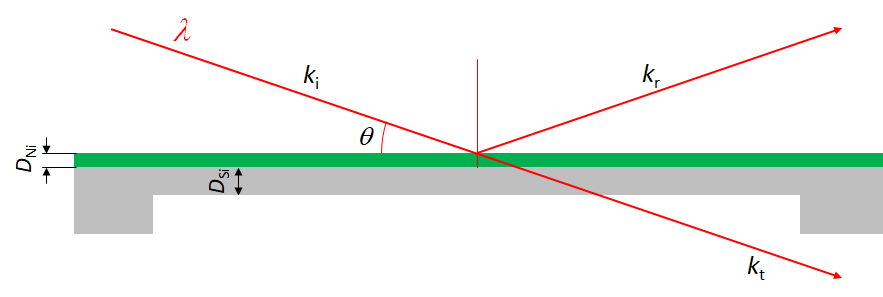}
	\caption{(Color online) The optical elements consist of a homogeneous Ni coating deposited on a Si substrate. The transmitted rays pass through both the thin Ni layer and the Si substrate, whereas the reflected rays undergo total external reflection directly at the air-coating interface. For the BS and REC mirror, the area of the Si wafer where reflection takes place is thinned down to approximately 50 $\mu$m.}
	\label{fig:Reflection-Transmission-BP}
\end{figure}

To reduce absorption of the transmitted neutrons by the Si-substrate for the BS and REC mirrors, the area were the neutron beam hits the substrate is thinned down to $D_{Si} = 0.05$ mm. The thinned part is coated with Ni having a thickness of $D_{Ni} = 122$ \AA. This way, the reflectivity becomes 50.0\% for $\lambda = 40$ \AA. The transmission of the neutrons through the Si is with 97.7\% high. As discussed above, the static attenuation is factored out by the magnetic lock-in extraction.

For the reflection mirrors $C_I$ and $C_{II}$, the Si substrates with a thickness $D_{Si} \simeq 0.7$ mm are coated with a Ni-layer with a thickness of $D_{Ni} = 1000$ \AA. The reflectivity of these mirrors including absorption is with $R = 99.9\%$ very high and can be replaced by $R = 1$.

\end{widetext}

\end{document}